# A global evidence map of human well-being and biodiversity co-benefits and trade-offs of natural climate solutions

Charlotte H. Chang[1,*]    James T. Erbaugh[2,3*]    Paola Fajardo[4,*]
Luci Lu[4,*]    István Molnár[6,*]    Dávid Papp[5,6,*]    Brian E. Robinson[4,*]
Kemen Austin[7]    Miguel Castro[8]    Samantha H. Cheng[9]
Susan Cook-Patton[8]    Peter W. Ellis[8]    Teevrat Garg[10]
Timm Kroeger[2]    Rob I. McDonald[11]    Erin E. Poor[8]
Lindsey Smart[8]    Andrew R. Tilman[12]    Preston Welker[8]
Stephen A. Wood[13]    Yuta J. Masuda[14,*,@]

[1] Department of Biology and Environmental Analysis Program, Pomona College, USA
[2] Global Science, The Nature Conservancy, USA
[3] Department of Environmental Studies, Dartmouth College, USA
[4] Department of Geography, McGill University, Canada
[5] Department of Telecommunications and Media Informatics, Budapest University of Technology and Economics, Hungary
[6] Lexunit, Hungary
[7] Forest and Climate Program, Wildlife Conservation Society, USA
[8] Tackle Climate Change, The Nature Conservancy, USA
[9] Global Science, World Wildlife Fund, USA
[10] School of Global Policy & Strategy, University of California San Diego, USA
[11] Center for Sustainability Science, The Nature Conservancy, Germany
[12] Northern Research Station, USDA Forest Service, USA
[13] Provide Food and Water, The Nature Conservancy, USA
[14] Partnerships and Programs, Vulcan LLC, USA
* Co-lead author
@ Corresponding author

# Abstract

Natural climate solutions (NCS) are critical for mitigating climate change through ecosystem-based carbon removal and emissions reductions. NCS implementation can also generate biodiversity and human well-being co-benefits and trade-offs ("NCS co-impacts"), but the volume of evidence on NCS co-impacts has grown rapidly across disciplines, is poorly understood, and remains to be systematically collated and synthesized. A global evidence map of NCS co-impacts would overcome key barriers to NCS implementation by providing relevant information on co-benefits and trade-offs where carbon mitigation potential alone does not justify NCS projects. We employ large language models to assess over two million articles, finding 257,266 relevant articles on NCS co-impacts. We analyze this large and dispersed body of literature using innovative machine learning methods to extract relevant data (e.g., study location, species, and other key variables), and create a global evidence map on NCS co-impacts. Evidence on NCS co-impacts has grown approximately ten-fold in three decades, although some of the most abundant evidence is associated with pathways that have less mitigation potential. We find that studies often examine multiple NCS pathways, indicating natural NCS pathway complements, and each NCS is often associated with two or more coimpacts. Finally, NCS co-impacts evidence and priority areas for NCS are often mismatched–some countries with high mitigation potential from NCS have few published studies on the broader co-impacts of NCS implementation. Our work advances and makes available novel methods and systematic and representative data of NCS co-impacts studies, thus providing timely insights to inform NCS research and action globally.

# Introduction

Natural climate solutions (NCS) are a subset of nature-based solutions consisting of 22 intentional actions ("pathways") to protect, restore, and manage forests, wetlands, grasslands, coastal systems, and agricultural lands to mitigate climate change (*1*). They are a necessary complement to rapid reductions in fossil fuel emissions (*2*), with nearly 96% of updated nationally determined contributions to the Paris Agreement including NCS (*3*, *4*). Enthusiasm for NCS is often grounded in their potential to advance development, conservation, and sustainability goals through the co-occurrence of biodiversity and human well-being benefits (i.e., "NCS co-benefits", or any combination of one or more NCS pathways and a human well-being or biodiversity co-benefit). Indeed, co-benefits are a consistent theme motivating NCS in scientific studies (*1*, *5–9*), policy reports and government documents (*10–12*), as well as broad appeals to accelerate NCS implementation (*13–16*). Assuming NCS yield co-benefits, they also bridge the Sustainable Development Goals (SDGs) (*17*), Paris Climate Accord (*18*), and Global Biodiversity Framework (GBF) (*19*). Recent high-level reports have noted the synergies between biodiversity conservation and climate change mitigation via NCS (*20*, *21*). Perhaps most importantly, NCS co-benefits have the potential to accelerate climate action by aligning local incentives through direct, near-term benefits to the communities who are often responsible for mobilizing NCS actions that ultimately provide a global public good (*22*, *23*).

Practical and political urgency underscore the need to understand if, how, and where NCS can deliver co-benefits. Only seven years remain to achieve the SDGs, accelerate action for ambitious climate pledges (*18*), and meaningfully advance recent commitments for land and biodiversity conservation (*19*). These global goals increase the financial and political capital that can be leveraged to make rapid and substantial investments to implement NCS at scale (*24*, *25*). While proponents of NCS emphasize that they can advance progress toward multiple goals, without a full accounting of the universe of evidence on NCS co-benefits, there are serious threats to the credibility of such claims. Further, in practice, decision-makers implementing NCS projects frequently grapple with trade-offs (*26*, *27*), and systematic evidence on both cobenefits and trade-offs remains poorly understood. Advancing NCS implementation without informed actions can threaten the enthusiasm and momentum seen in recent years, and may even lead to unanticipated adverse impacts on people and nature.

Despite the urgent need for a global stocktake on NCS co-benefits and trade-offs (i.e., "NCS co-impacts"), we are unaware of any effort to systematically analyze and map the broad universe of NCS co-impacts studies. To date, studies have provided circumstantial assessments of NCS co-benefits (e.g., *1*, *5*), or conducted systematic assessments of one of the 22 NCS pathways, usually in isolation (e.g., *28–34*). Nearly all studies acknowledge the need for a consistent systematic evidence map on all NCS pathways, co-benefits, and trade-offs. A global evidence map on NCS co-impacts is critical for broad and strategic adoption of NCS because it clarifies the full value of NCS through the value they bring via mitigation and co-benefits. Such an evidence map would allow communities, decision-makers, investors, and others with vested interests to evaluate the costs and full benefits provided by NCS, assessing potential trade-offs between multiple objectives, and deploying resources more effectively.

We identify and map evidence from peer-reviewed publications on the relationship between NCS co-impacts. We overcome two significant challenges that have so far hindered systematic mapping of NCS co-impacts evidence. First, research on NCS spans many fields, such as animal science, ecology, economics, public health, soil science, agronomy, traditional ecological knowledge, toxicology, and environmental governance. Much of the evidence long predates the organizing concept of "NCS," representing a diverse set of potential evidence that can be challenging to consistently characterize (*35*, *36*). Searching and identifying relevant evidence across a diversity of disciplines is practically challenging due to divergent or siloed epistemologies and semantic ontologies (*37*); thus, identifying the relevant topic space is methodologically vexing (*38*). Second, because cross-cutting topics often generate large volumes of potentially relevant articles, previous efforts that have examined such topics have relied on hundreds of researchers who spend thousands of hours manually searching, screening, and coding papers (e.g., *39–41*). Advances in machine learning ("ML") assisted methods have led to greater efficiencies, but these efforts often still require significant financial and human resources (*39, 40*) that lead to evidence maps that are often several years out of date by the time they are completed and published. Even with efficiency gains from ML assisted methods, researchers have still needed to constrain the universe of eligible studies to generate a subset of manuscripts manageable for human review (*40, 42*). Indeed, other efforts to review similar literature have limited search results to ensure "manageability" (e.g., *43*), but this can artificially limit the scope of the literature being surveyed. Rapid advances in ML for natural language processing ("NLP") have yet to be fully integrated into systematic reviews, but studies such as Callaghan et al. (*41*) show the benefits of adopting a ML and expert human workflow for rapid evidence synthesis.

Our evidence map overcomes barriers to scale by using a novel application of large language models to categorize abstracts, thereby identifying bodies of research on NCS and their coimpacts (Methods). Our search strategy is deliberately inclusive of co-benefits and trade-offs, resulting in an evidence map that captures the full distribution of co-impact evidence measured by the volume of publications. This analytic choice derives from our recognition that implementing NCS may come with local trade-offs in human well-being, biodiversity, and climate change mitigation outcomes (*26, 27*). We also use state-of-the-art text parsing algorithms (e.g., *44, 45*) to extract information on study geography, biodiversity (species), and cost information, in addition to identifying abstracts with content related to Indigenous Peoples and Local Communities (IPLCs) and equity considerations (see SI for detailed methods). The result is a replicable evidence map pipeline that can be easily and regularly reproduced.

We present an evidence map of 257,266 relevant papers on 22 NCS pathways and 11 coimpacts (Methods for details and Extended Data Tables 1-2 for definitions) from an original sample of 2.28 million unique papers (fig. 1). Our evidence map consists of English language papers from 1990-2022 (fig. 2). The papers in our evidence map represent 181 countries, all five biomes, 246 disciplines, and 364 unique thematic topics (details on topics in fig. S3 and Extended Data Table 7). Following the NCS hierarchy (*6*), we group NCS into broader categories of protect, manage, and restore actions. The vast majority of papers (87%) included manage pathways (e.g., conservation agriculture, trees in croplands, natural forest management), while 30% of the papers covered protection (e.g., avoided forest, wetland, grassland conversion), and 29% covered restoration pathways (e.g., wetland, grassland, and forest restoration). Many of the papers contained information on biodiversity (41%) or biome (39%), but some attributes were relatively rare: we detected that fewer than 2% of the papers contained cost, equity, or IPLC-related information (see Extended Data Tables 3-6 for search queries). Our search results comprehensively encompass other NCS co-impact evidence mapping efforts that have had a constrained and expert-reviewed sample (*42*, *46*, *47*, table S2).

Our resulting evidence map is a comprehensive dataset of relevant papers that provides insights on evidence gaps (where more research is needed) and high-priority areas for further investigation and investment (where abundant evidence exists). This map provides foundational evidence needed to accelerate strategic NCS research and implementation for climate change mitigation and adaptation, biodiversity conservation, and sustainable development.

[Figure 1 about here.]

[Figure 2 about here.]

[Figure 3 about here.]

## Results

### NCS evidence base, gaps, and action areas

Our evidence map highlights two areas of evidence needs based on (a) the overall distribution (fig. 3) and (b) the spatial breakdown of NCS co-impacts evidence (fig. 4). Looking across the overall distribution of evidence (fig. 3), many of the NCS pathways with the highest carbon mitigation potential have the highest level of evidence and vice versa. Exceptions to this pattern include wetland protection and restoration, a pathway that includes important habitats such as peatlands, which can contribute approximately 6% of total global NCS mitigation potential by 2030, respectively (*1*). Cumulatively, there is a greater concentration of evidence in management-related NCS compared to protection (1.8 times more evidence per-NCS pathway) and restoration (3.3 times more) pathways. While the global body of evidence on NCS coimpacts has grown in total volume over time, the growth is far from uniform (fig. 3). We plotted normalized deviates for literature growth for each NCS–co-impact combination (the NCS-coimpact pair minus the overall trend in the data), and observed that combinations such as avoided forest conversion and economic living standards, or grazing (legumes in pastures) and education exhibited sustained growth, while others such as cropland nutrient management and human health, or biochar and biodiversity represent a decreasing share of published NCS co-impact studies.

The evidence map, however, is not necessarily correlated with mitigation potential. For instance, conservation agriculture has the most evidence of any pathway, but its maximum mitigation potential of 516 Tg $CO_2$ per year by 2030 is less than that of biochar, trees in croplands, or optimizing grazing intensity. Conversely, a lack of evidence (empty cells) in fig. 3 does not necessarily mean there is no meaningful NCS-impact link. For example, we may have limited evidence for grassland restoration and education or social relations, but the increased uptake of NCS makes it critical we generate better evidence on the possible co-benefits and trade-offs.

We compare regions based on their co-impact evidence base and their climate change mitigation potential to determine which areas may be prioritized for NCS action and where there may be an evidence base to inform whether NCS can help address human or environmental challenges. We merge geolocated articles from our database with climate mitigation potential from NCS (*1*, *5*, *48–54*), threatened biodiversity (*55*), and human well-being (*56*) at the country level. We stratify the articles by NCS pathways using protect, manage, and restore categories (*6*). By combining these datasets, we identify regions with high NCS potential and large bodies of evidence ("action areas") versus areas that have high NCS potential but relatively little evidence ("need areas") (fig. 4).

We observe direct relationships between NCS mitigation potential and publications on NCS co-impacts for the protect, manage, and restore pathways (fig. 4; (*1*, *5*, *48–54*)). This relationship is strongest for NCS management pathways ($\beta = 0.77, p < 0.001$), and is weakest for protection ($\beta = 0.31, p < 0.001$) and restoration pathways ($\beta = 0.41, p < 0.001$). Thus, many countries with the greatest potential to contribute to climate mitigation through the protection or restoration of natural habitats contain comparatively fewer studies on the biodiversity, human well-being, or environmental benefits or trade-offs that NCS can provide.

High-income countries and countries with the greatest NCS mitigation potential constitute the preponderance of evidence on NCS co-impacts. For example, Australia, Brazil, Canada, China, India, Indonesia, Mexico, and the United States are action areas across all protect, manage, and restore pathways. The comparatively large amount of peer-reviewed publications that focus on NCS co-impacts in these countries, combined with their overall potential for climate mitigation through NCS, reinforce the importance of leveraging current findings to rapidly inform where and how to implement NCS that may also deliver well-studied benefits (fig. 4).

A number of countries with great mitigation potential but comparatively little published evidence constitute "need areas." Regions with multiple countries categorized as need areas in at least two of the three pathway categories ($n$ = 17) include Central South America (Paraguay and Uruguay), West and Central Africa (Ivory Coast, Gabon, Guinea, Mali, the Republic of Congo, and Sierra Leone), and Central Asia (Kazakhstan and Uzbekistan). Ranking "need areas" according to their human development index (HDI) and threatened species richness indicates that West and Central African nations comprise six of the top ten countries with the highest combined number of threatened species and lowest HDI. Together, countries categorized as need areas for one or more of the pathway categories contain over 124 million people who are multi-dimensionally poor (*57*) and 2,162 endemic species of animals and plants (*55*).

Stakeholders may vary in how they weigh climate mitigation versus the co-impacts associated with NCS implementation. Equal weighting for HDI, threatened species richness, and climate mitigation from NCS yields a similar set of areas with fewer publications on NCS co-impacts across all pathway categories. These include Sierra Leone, New Caledonia, the Republic of Congo, Burundi, Mauritius, Somalia, the Solomon Islands, Haiti, Puerto Rico, and Yemen, among others (table S4).

[Figure 4 about here.]

## Co-occurrence of NCS and co-benefits

We find NCS pathways are rarely studied in isolation. 70% of papers (n=179,876) are predicted to contain information on more than one NCS pathway (fig. 2), and 29% of papers ($n$ =74,608) on more than three pathways. Pathways that most commonly co-occur within the data include nutrient management and conservation agriculture ($n$ =33,167) and natural forest management and avoided forest conversion ($n$ =31,745). Examining the most significant trends for individual pathways, over 96% of papers on fire management also consider natural forest management ($n$ =17,232), and 61% of papers on avoided coastal wetland impacts and conversion co-occur with research on coastal wetland restoration ($n$ = 20,814). Though the most common and proportionately significant co-occurrence exists within anthromes, some pathways also show considerable co-occurrence across anthromes. For example, 14% of papers that examine conservation agriculture also consider natural forest management ($n$ =13,955), and 12% include avoided forest conversion ($n$ =10,899). Although we cannot distinguish whether there is complementarity or trade-offs among these pathways, the prevalence of co-occurrence suggests that NCS research strives to take a landscape-level approach when investigating NCS outcomes.

Like the co-occurrence of NCS pathways, potential NCS co-impacts were often studied together. Most (94%) studies are predicted to analyze more than one potential co-impact ($n$ = 241,308). Economic, material, and environmental/ecosystem service outcomes demonstrate the greatest amount of co-occurrence. For example, 34% of studies that examine economic co-impacts also include information on potential material co-impacts ($n$ =114,270) and 28% include information on co-impacts for environmental/ecosystem services ($n$ = 95,622). The common co-occurrence of potential co-impacts, such as those involving human health and biodiversity ($n$ = 8904) and human health and environmental/ecosystem services ($n$ = 18,919), may reflect a growing awareness of the interplay between these sectors, evident also within literature on one or planetary health (*58*). Similarly, 15% of papers that include information on culture and spirituality also contain analyses of economic impacts ($n$ = 14,210), which may reflect a concern that economic development will negatively impact local traditions. Further research will be necessary to determine the direction of this and other relationships among potential co-impacts.

[Figure 5 about here.]

## Discussion

The IPBES-IPCC report (*20*) articulates the critical importance of simultaneously advancing climate change mitigation, biodiversity conservation, and sustainable development objectives. NCS are climate-focused nature-based solutions (NBS) that have gained significant interest and investment in multilateral frameworks and national policies to address multiple societal challenges (*59*). Yet, there are still significant uncertainties on whether and where NCS can deliver on multiple objectives, as well as the potential trade-offs that must be considered (*20*). Using established frameworks on NCS pathways (*1*) and co-impacts (*60*), our evidence map provides, for the first time, a global stocktake on co-impacts provided by all 22 NCS pathways, thus providing a foundation to guide actions and research priorities for all NCS pathways.

The evidence map presents the latest distribution of evidence across the 22 NCS pathways, including where there has been relatively abundant attention and research (e.g., forest protection; North America) and a more recent focus (e.g., coastal restoration; parts of West Africa). Importantly, our evidence map also shows that we have ample evidence for many of the NCS pathways with high climate change mitigation potential, but there are exceptions such as wetlands protection and restoration pathways, despite their estimated 6% contribution to climate change mitigation potential, globally (*1*). While the total volume of evidence about the potential co-impacts of NCS has grown over time, the growth is not uniform. Evidence map users can conduct deeper dive investigations along the various facets that are available in our dataset, such as by pathway, geography, co-impact type, and biome. To date, the global distribution of evidence on NCS co-impacts has been unknown, and a clear next step is a careful assessment of the direction (positive, negative, or neutral) and size of impacts for a given geography and sub-population of interest for any given NCS co-impact. Like other evidence maps that conducted global stocktakes (e.g., *39*, *41*), we did not assess these aspects given the sheer volume of evidence.

Analyzing the overlap between areas of high or low evidence with indicators for NCS climate mitigation potential, threatened biodiversity, and human development provides deeper insights into where more research is needed versus where abundant evidence exists to inform action. Brazil, India, China, Mexico, the United States, Australia, and Canada are high-mitigation potential countries with abundant co-impact evidence. On the other hand, there are countries, predominantly in West and Central Africa, which lack extensive published evidence despite high mitigation potential, threatened biodiversity, and opportunities for improving well-being. For stakeholders who equally value carbon alongside biodiversity and potential social benefits, a different set of priority countries emerge such as New Caledonia, Burundi, Mauritius, Somalia, Haiti, and Yemen. It is possible that these countries have little evidence in our dataset because we focused on articles published in English or historical biases in science funding and capacity; however, the spatial distribution of our evidence map mirrors that of other reviews (*39*, *60*).

We also found that the majority of management NCS pathways co-occur, indicating that implementation around just one pathway in a particular location may overlook the complementarity of some NCS pathways that have been studied together in the literature. Conservation agriculture, nutrient management, and trees in croplands are likely to be compatible activities, and programs focused on a singular NCS pathway may miss key opportunities for climate change mitigation and the multitude of co-impacts that may emerge from a bundled approach. However, each NCS pathway likely has unique challenges around planning and implementation, so program managers should be mindful of how constraints around NCS implementation can vary (e.g., *61*). Given that the funding gap to reverse biodiversity decline alone by 2030 is between USD 722-967 billion per year (*62*), and that nearly USD 400 billion per year is needed to implement forest-based NCS alone (*63*), identifying synergistic opportunities where multiple NCS can cost-effectively advance multiple goals will be important in a resource constrained world. The portfolios of co-impacts highlighted by our evidence map provide a blueprint for the possible architecture of climate action building blocks (*23*).

Examining NCS co-impacts has long been hampered by the fact that any analyses must sift through a large body of existing literature across disciplinary boundaries and determine which studies contain information on NCS. Research about the impacts of protection, improved management, and restoration of ecosystems has existed for decades. In most cases, this research was conducted before "NCS" proliferated as a climate change mitigation strategy and term, necessitating searching for individual pathway names and pseudonyms (e.g., reforestation, forest restoration) rather than using "natural climate solution" as a catchall. A comprehensive analysis thus necessitates analyzing a scale of published studies that defies manual review. We are unaware of any data-driven taxonomy that could identify which articles and topics map to NCS pathways prior to our study. Our unsupervised large language topic model permitted us to discover a categorization of literature to NCS pathways and co-impacts using established frameworks (*1*, *60*). Before recent watershed advances in language modeling, processing the relatively nuanced differences between scientific abstracts would have been difficult, if not impossible, with existing models. An important advance is that our evidence map pipeline is replicable and easily updated given our use of modern ML approaches. Such built-in automation decreases effort and costs significantly while preserving scientific robustness.

Nevertheless, there are ample opportunities for future advances. The most notable is how we predict relevant articles – not every study in our sample may fulfill more stringent inclusion criteria, such as an estimation of an effect size or direction of impact. However, our dataset allows a drastic reduction in search and screening effort by providing a funnel to direct more targeted meta-analyses. At a minimum, our approach systematically and rigorously identifies current research effort and gaps for future research. While computer vision approaches that extract tables could offer additional machine-aided steps, we are unaware of any way that more intensive analyses such as a meta-analysis could be reliably performed end-to-end by a machine learning pipeline, especially across such a diverse body of research. Our analysis was also limited to papers written in English, which may be contributing to the relative lack of evidence in some geographies, such as West and Central Africa. Expanding the machine-aided pipeline to capture non-English papers will be critical to ensure equitable evaluation of all scientific evidence that is generated across diverse geographies. Further, if there were relatively small and isolated bodies of literature that could be pertinent to NCS co-impacts, our large language topic model may have failed to identify these clusters of papers.

Strategic and targeted implementation of NCS is an important component for realizing environmental, biodiversity, and human well-being goals, as well as for assessing trade-offs in the event that achieving multiple aims is challenging (*20*). Enthusiasm for NCS has, in part, been rooted in the assumption that they will deliver on multiple objectives (*1*, *5*, *6*, *10–15*), yet significant barriers have hindered systematic evidence maps that could support such assertions. An important next step will be to assess the direction, size, distribution, sustainability, and timing of co-impacts offered by NCS pathways in isolation or combination, and potential trade-offs that implementation of NCS may present, especially for vulnerable or already marginalized populations and endangered species (*64*). Given the increased political will and financial commitments toward climate change mitigation (*24*, *25*), and the relative immediacy and readiness of NCS implementation compared to other forms of carbon mitigation, mapping where and what co-benefits and trade-offs NCS realizes can provide valuable and actionable insights to inform global climate change mitigation that can also address biodiversity conservation and human development.

## Methods

Our analysis involved five major steps (Figure 1 and fig. S1): 1) developing search strings corresponding to the NCS pathways and co-impacts framework to query academic research databases; 2) filtering articles based on distance to each NCS pathway's centroid; 3) training an unsupervised machine learning topic model ("BERTopic") to identify categories that were then mapped to NCS pathways and co-impact categories; 4) geolocating abstracts and identifying the biome associated with specific locations using machine learning methods (e.g., named entity recognition used in packages such as Mordecai (*65*)); 5) determining which species were mentioned in the abstracts. For more detailed information regarding any of these steps, please refer to the Supplementary Materials. We conducted several robustness checks with human review and compared against other evidence mapping efforts (details in SI Section 3.2)

### Compiling NCS evidence

We performed 242 queries to Web of Science and Scopus in August 2022 resulting in 2.28 million unique citations, or abstracts with metadata. The search strings were informed by IUCN definitions of biomes (*66*) and Griscom et al.'s (*1*) definition of NCS pathways. We used McKinnon et al.'s (*60*) framework for human well-being categories. Our search strings comprehensively captured the presently known universe of NCS co-impacts; our search results contained over 96% papers captured in other evidence map or review projects ((*42*, *46*, *47*), SI Section 3.4). For each search string query, we defined a more restrictive set of terms ("focused queries") designed to yield a higher proportion of pertinent abstracts and a less restrictive set of terms ("broad queries") that would sample the broader universe associated with each search target, that is, an NCS pathway and a type of co-impact. The team iterated and calibrated search strings until the results were deemed by team members to be inclusive of relevant papers on a random selection of articles. The focused queries yielded 800,000 abstracts while the broad queries produced 1.4 million additional abstracts. Each abstract was converted to a numerical representation (an "embedding") using a large language transformer model, specifically, SentenceBERT (*67*).

Categorizing and identifying relevant abstracts

We used cosine distance to the centroid of the specific query abstracts for each pathway to threshold which of the broad query abstracts were retained for additional analyses (SI Section 3). Based on the thresholding, we were then left with 1.28 million abstracts. At this step, our sample contained 93.6-99.8% of the papers analyzed in manually-reviewed NCS evidence mapping exercises (*42*, *46*, *47*, table S2). We then applied BERTopic, an unsupervised learning approach that builds upon traditional topic modeling using simpler bag-of-words features with cutting-edge large language model, transformer-based embeddings that capture the semantic structure of language (*68*).

BERTopic has been used to describe viewpoints toward climate change belief (*69*) and to perform a systematic review of machine learning in urban studies (*70*). The major advantage of using BERTopic is that it can scale to large text datasets, such as ours, and does not require a preexisting labeled dataset, which itself necessitates a known taxonomy of NCS pathways and co-impacts (*70*). At the outset of our research, we were uncertain which, if any, of the NCS pathway and co-impact combinations would have any coverage in academic research, and thus sought an approach that would allow categories to emerge from the data.

We performed hyperparameter optimization on the BERTopic model. The final model generated 800 topics which the author team inspected for relevance, defined as having a clear connection to NCS pathways and at least one type of human well-being or environmental coimpact. We also coded topics for NCS pathways and co-impacts. At all steps, at least two coders examined a topic for its relevance, NCS pathway, and co-benefit. At this juncture, our dataset contained 70.4-77.1% of the NCS biodiversity co-impact papers analyzed by expert teams (*46*, *47*, table S2). The team discussed any disagreements in topic coding and resolved any differences between coders. We dropped topics that did not capture an NCS pathway and co-benefit (436 topics dropped), resulting in a total of 364 topics covering 257,266 articles.

## Extracting variables from abstracts

We used Python modules to extract cost information, geolocate abstracts, and identify taxa mentioned in abstracts. The cost information and geolocation approaches used named entity recognition approaches, while the biodiversity extraction used regular expressions for the scientific and common names of species drawn from the Open Tree of Life (*71*). For more information, please refer to SI Section 4.

## Analysis

We performed descriptive analyses of trends in evidence, evidence gaps, and NCS co-impact co-occurrence. For NCS evidence gaps focused on identifying high priority areas for action and need, we combined information on country-level mitigation potential for protect, manage, and restore pathways (*6*), threatened biodiversity (*55*), and human development (*56*). All analyses were conducted using Python (v. 3.11) or R (v 4.1), and data and code are available upon reasonable request.

# Acknowledgements

We are grateful to Jonathan Levine, Will Petry, Bronson Griscom, Frances Seymour, Cecile Girardin, Lydia Olander, and several others for their thoughtful feedback, as well as attendees at various presentations which strengthened our analysis and manuscript. We thank Daniel Lovas´ for technical contributions that greatly enhanced the manuscript. The findings and conclusions in this publication are those of the authors and should not be construed to represent any official USDA or U.S. Government determination or policy. Funding from the Bezos Earth Fund and other donors supporting The Nature Conservancy benefited this project.

# Contributions

CHC, BER, JTE, LL, YJM, IM, DP, and PF conceived the idea for the manuscript, generated new methods, and performed the research. CHC, BER, JTE, YJ, IM, DP, and SCP performed data analysis and modeling, generated figures and tables, and drafted sections of the manuscript. ART, SCP, RIM, TG, PWE, EP, TK, SHC, PW, SAW, MC, LS, KA, and PF performed critical reviews. YJM, EP, JTE, and RIM obtained funding.

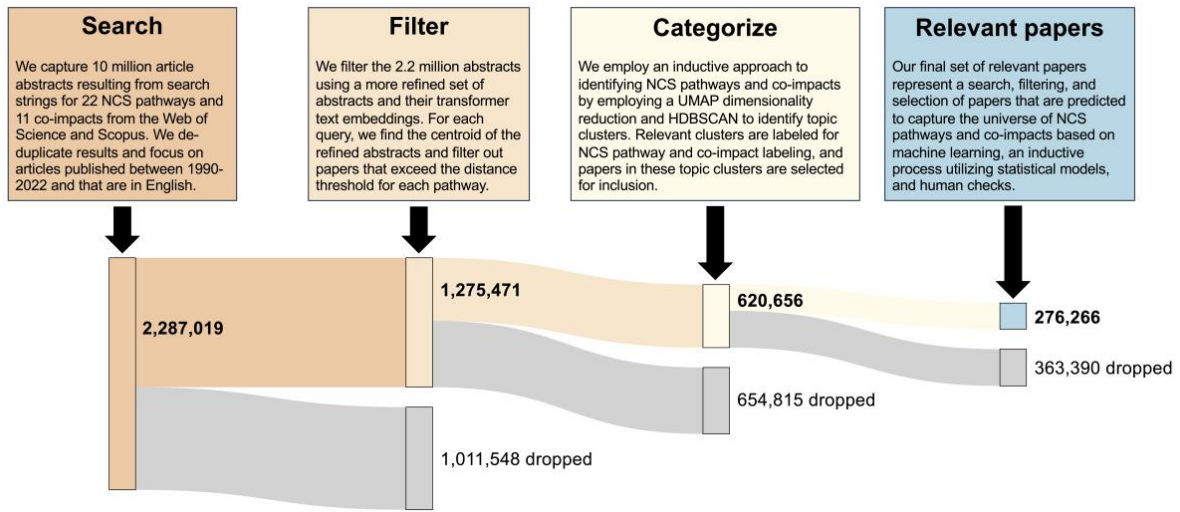

Figure 1: Evidence base generation process. *The data generation pipeline broadly involves four steps where we employ large language models to categorize abstracts to identify bodies of research on NCS co-impacts, state-of-the-art text parsing algorithms to extract information, alongside human review for check for robustness. Details of each step are outlined in the Methods and Supplementary Materials.*

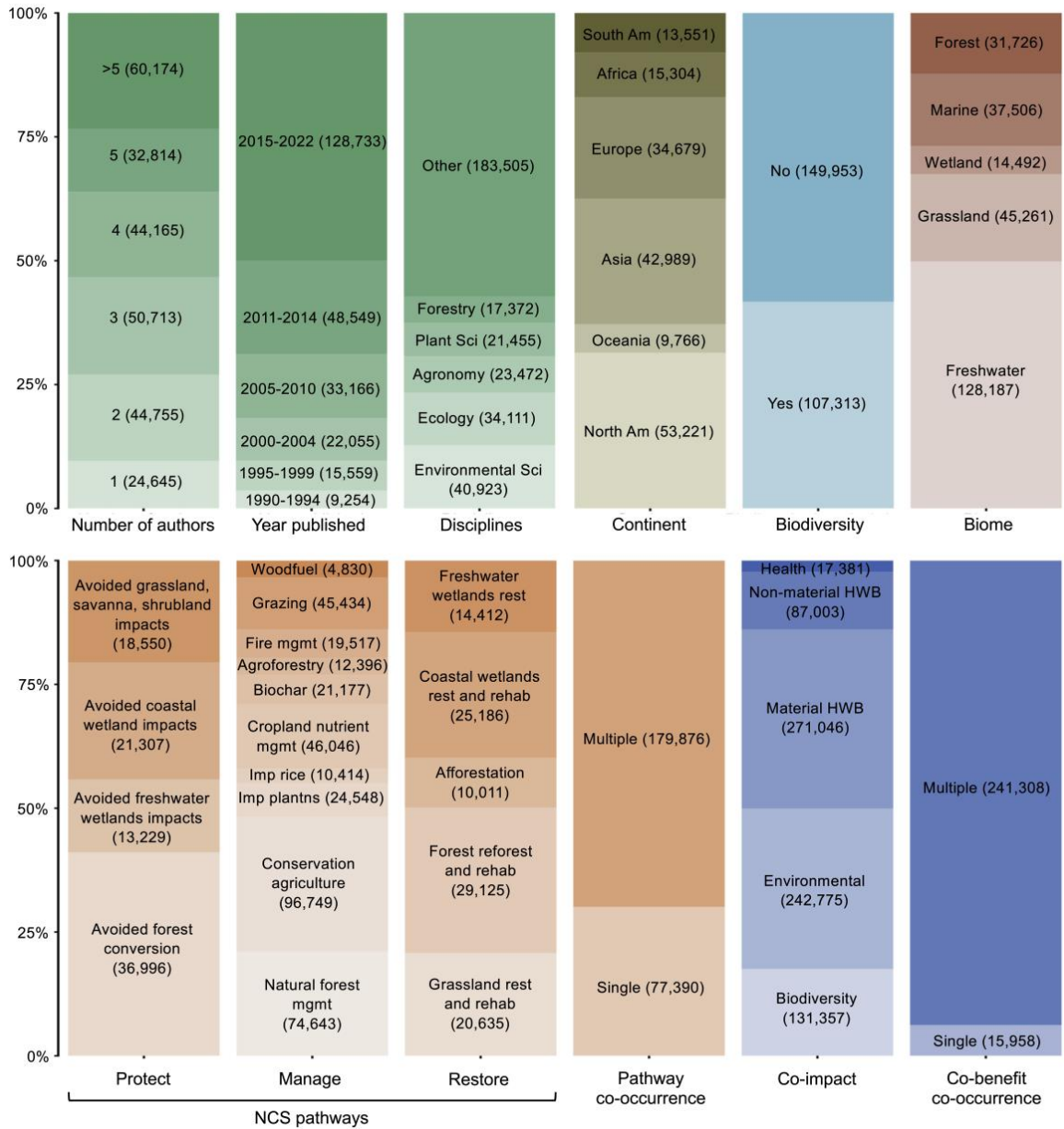

Figure 2: Characteristics of the evidence base. *Each category is normalized to 100% of the relevant papers, where we do not display papers with missing data for disciplines, continent, and biome. Further information on the Other category for Disciplines is available in Extended Data Table 9. Data for various Grazing pathways are not shown due to space limitations. Grazing categories include optimal intensity, legumes in pastures, improved feed, and animal management include 8,274, 4,580, 10,469, and 22,111 articles, respectively. Some terms are shortened or abbreviated for space, such as human well-being (HWB) and plantations (plantns).*

*Descriptions for NCS pathways and co-impacts are shown in Extended Data Tables 1-2.*

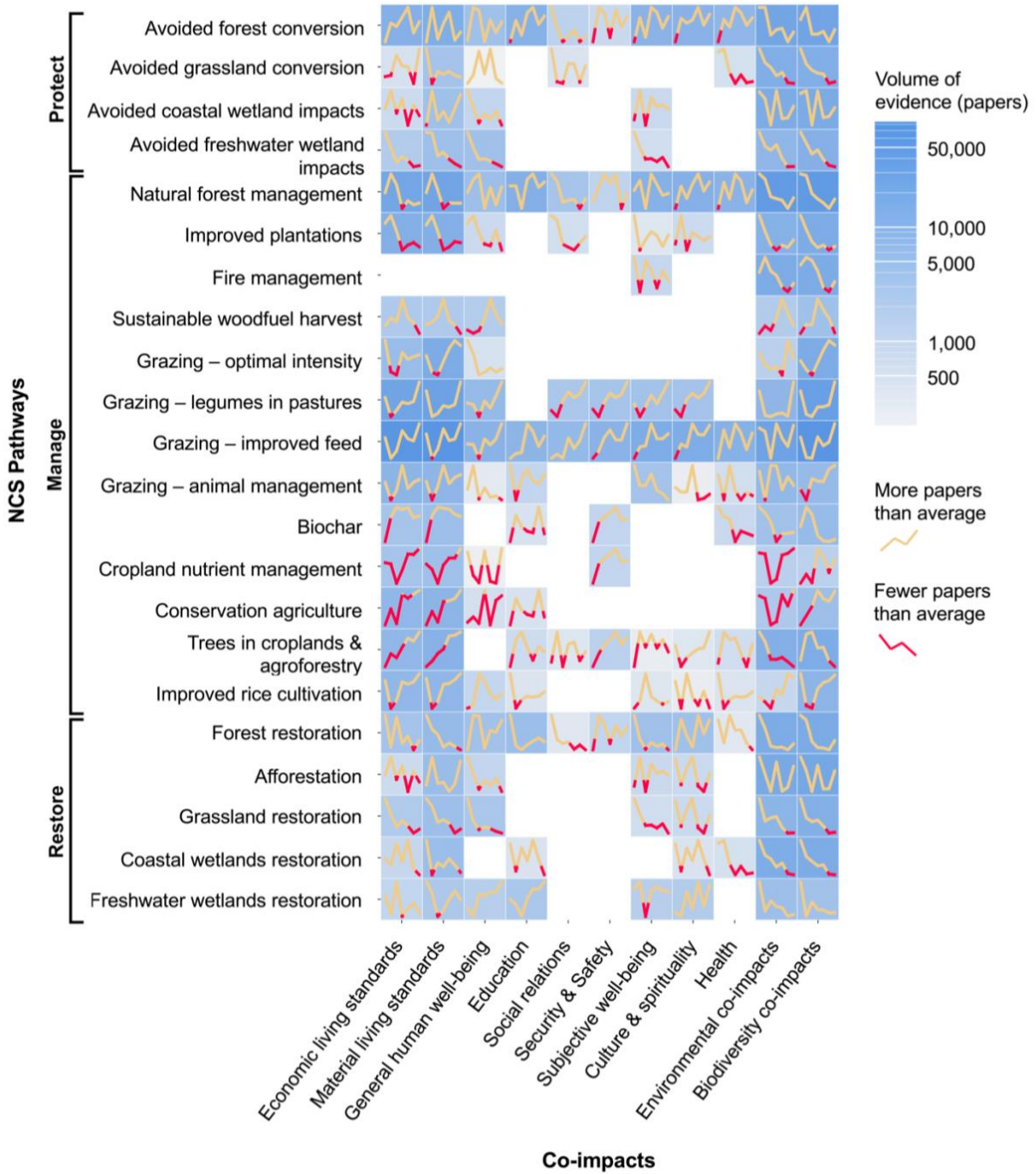

Figure 3: NCS and co-benefit evidence base. *Darker shaded cells denote greater than average volume of papers relative to the total evidence base. Line graphs within each cell represent how evidence for the NCS co-impact combination changed over time. Yellow lines indicate year-over-year growth that exceeds the average growth of the overall body of evidence; red lines*

*indicate lower than average growth of NCS papers discussing this pathway-co-impact combination. Definitions for NCS pathways and co-impacts are provided in Extended Data Tables 1-2.*

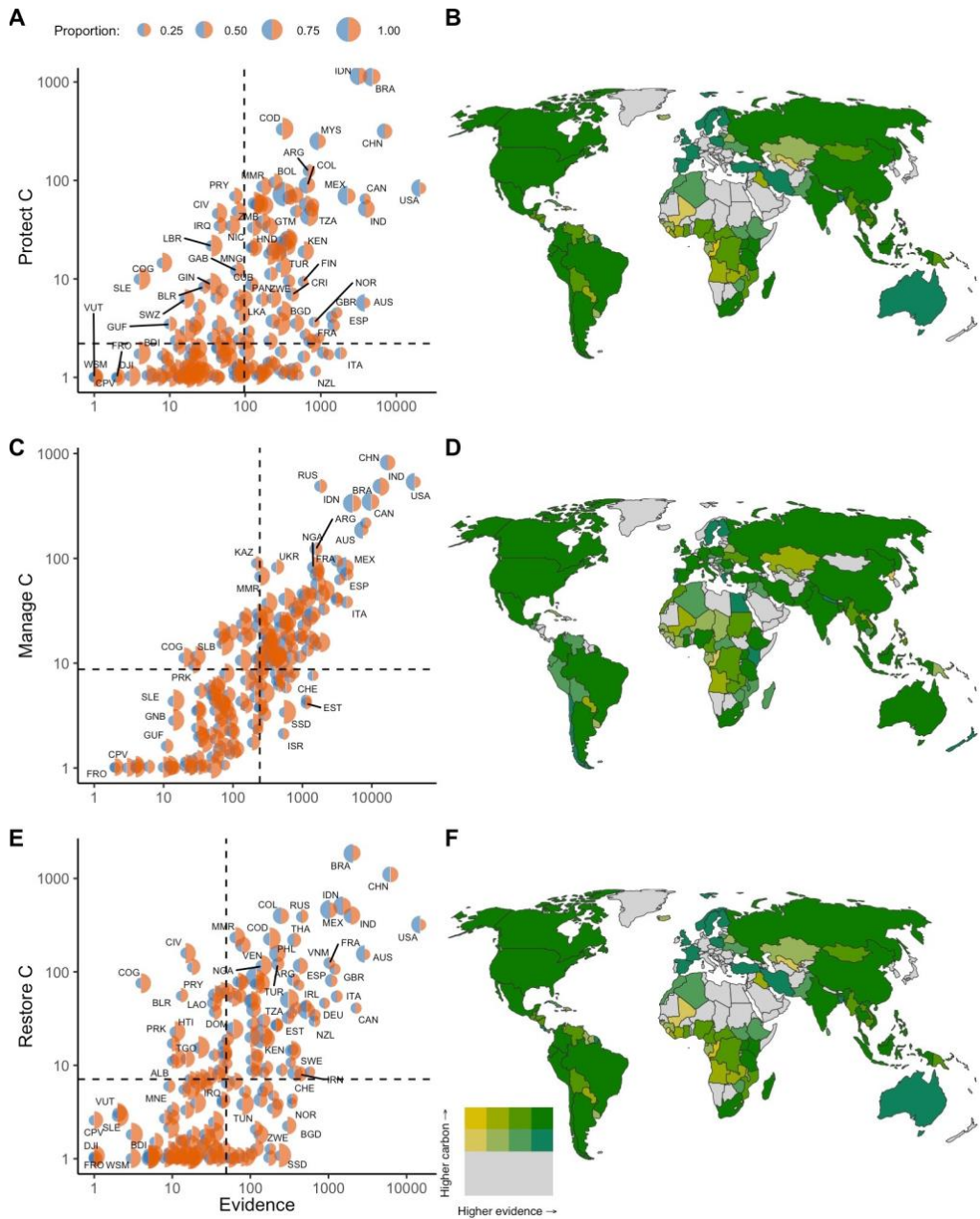

Figure 4: The volume of evidence for NCS strategies across countries versus their maximum climate mitigation potential through NCS and threatened biodiversity. The maximum value for threatened biodiversity is 3,747 species while for HDI it is a minimum of 0.361. *A, C and E show semi-circles on the left (blue) for threatened biodiversity and on the right (orange) for the national human development index. Larger semi-circles correspond to a country having higher threatened biodiversity or lower human development. The dashed vertical line corresponds to the country-median quantity of evidence, while the dashed horizontal line marks the country median climate mitigation value for each strategy. B, D and F show maps colored based on the*

*degree of climate mitigation potential for each pathway as well as the paucity of evidence. Countries that are more intensely colored in gold are those that have high climate mitigation potential but limited evidence. Details of country-level data provided in table S4.*

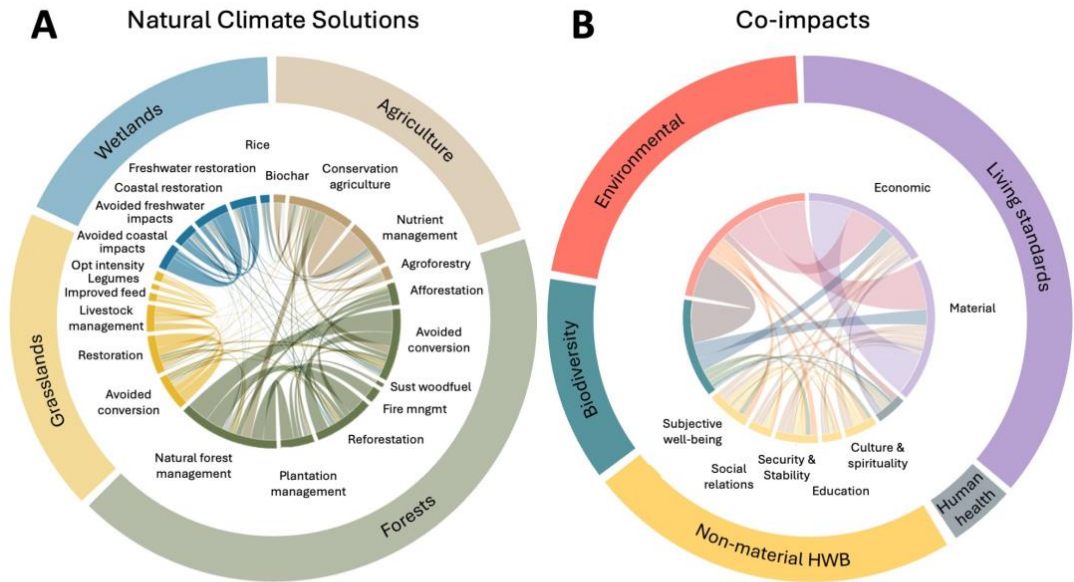

Figure 5: Proportion of pathway (A) and co-impact (B) co-occurrence. *Articles that have more than one NCS pathway (n = 179,846) or co-impact (n = 241,308) are displayed. Articles that contain more than two pathways or co-impacts are represented by multiple dyadic chords. Detailed information on co-occurrence provided in Extended Data Table 7.*